# Decade-long periodicity study of 2FHL blazars with historical optical data

Sagar Adhikari ,[1]★ Pablo Peñil ,[1] Alberto Domínguez ,[2] Marco Ajello ,[1] Sara Buson [3,4] and Alba Rico [1]

[1]*Department of Physics and Astronomy, Clemson University, Kinard Lab of Physics, Clemson, SC 29634-0978, USA*  
[2]*IPARCOS and Department of EMFTEL, Universidad Complutense de Madrid, E-28040 Madrid, Spain*  
[3]*Julius-Maximilians-Universität Würzburg, Fakultät für Physik und Astronomie, Emil-Fischer-Str 31, D-97074 Würzburg, Germany*  
[4]*Deutsches Elektronen-Synchrotron DESY, Platanenallee 6, 15738 Zeuthen, Germany*

## ABSTRACT

In our recent investigation, we utilized a century's worth of archival optical data to search for a decade-long periodicity from the blazar PG 1553+113, finding a hint of a 22-yr period. Building on this foundation, the current study extends our analysis to include 10 blazars from the *Fermi*-Large Area Telescope Second Catalog of Hard Sources (2FHL) catalogue to uncover similar long-term periodic behaviour. To ensure the reliability of our findings, we consider the impact of observational limitations, such as temporal gaps and uneven sampling, which could potentially introduce artefacts or false periodic signals. Our analysis initially identifies decade-scale periodicity in four of these blazars (AP Librae, MKN 421, MKN 501, PG 1246+586). However, further investigation reveals that three of these are likely influenced by noise and poor sampling. The most promising candidate, approximately 51 ± 9 yr signal in MKN 421, corresponds to fewer than three full cycles and cannot be considered significant. Furthermore, global significance suggests none of the candidate periodicities meet the threshold for statistical significance. These results underscore the importance of accounting for sampling artefacts and highlight the need for robust methodologies in long-term periodicity searches.

**Key words:** methods: data analysis – astronomical data bases: miscellaneous – time – (galaxies:) BL Lacertae objects: general.

## 1 INTRODUCTION

Active galactic nuclei (AGNs) are a phenomenon in the nuclei of galaxies that produce a tremendous amount of energy powered by the gas accretion on to a supermassive black hole (SMBH) compared to that of the aggregate of stars in the galaxy. If an accreting SMBH launches its jet towards us from these nuclei, they are known as blazars (e.g. Peterson 1997; Wiita 2006).

Blazars show variability across the electromagnetic spectrum and at different time-scales (Urry & Padovani 1995; Acciari et al. 2011; Chakraborty et al. 2015; Arbet-Engels et al. 2021). Traditionally, temporal variability in blazar light curves (LCs) is thought to range from intraday (∼minutes – a day; Wagner & Witzel 1995) to long-term time-scales (∼months – years; Jurkevich, Usher & Shen 1971). Recently, Peñil et al. (2020) identified years long periodicity in γ-rays in some blazars. In addition to that, recent studies using historical optical data on OJ 287 and PG 1553+113 hint towards even longer term variability (Abdo et al. 2009; Dey et al. 2018; Peñil et al. 2024). Decade-long variabilities in blazars provide valuable information on AGNs' physical processes and emission mechanisms (e.g. Bhatta 2021; Gupta et al. 2022; Adhikari et al. 2024, hereafter A24; Peñil et al. 2024). The detection of decade-long quasi-periodic oscillations (QPOs) can favour certain emission mechanisms and scenarios, such as binary SMBH systems or jet precession, which can significantly impact our understanding of the AGN central engine and its environment.

Binary SMBH systems are some of the models that explain long-term periodicity in blazars. In these systems, the interaction between two black holes can lead to periodic changes in the accretion rate and jet properties, resulting in observable QPOs (e.g. Tavani et al. 2018; Qian et al. 2019). The periodic gravitational influence of the secondary black hole on the accretion disc of the primary can cause variations in the emission intensity. For example OJ 287 exhibits a ∼12-yr period associated with the binary orbit and a ∼60-yr period due to the precession of the orbit (Dey et al. 2018). In addition, accreting binaries often hollow out a cavity and cause an overdensity in the circumbinary disc. The overdensity, referred to as a 'lump,' can periodically modulate the accretion rate of the primary black hole, causing periodic modulation in the jet emission as explored for decade-long QPO of PG 1553+113 (A24).

Additionally, jet precession driven by the misalignment of the black hole's spin axis and the accretion disc can produce QPOs (e.g. Rieger 2004). The precession causes the jet direction to change periodically, altering the observed emission when the jet orientation relative to our line-of-sight changes. This model is supported by multiwavelength observations that show coordinated variability patterns over long time-scales (e.g. Qian et al. 2018). Therefore, investigating decade-long QPOs in blazars not only provides an understanding of the dynamics of SMBH systems but also enhances a broader comprehension of jet physics and AGN variability, contributing to our knowledge of cosmic evolution and the role of AGNs in shaping their host galaxies.

★ E-mail: sagara@clemson.edu

In A24, we analysed historical optical LC of PG 1553+113 and found hints of ∼22 yr periodicity along with a 2.2 yr period attributing the longer period to overdensity on the circumbinary disc. We also found that the period was unlikely to be an artefact of the sampling of data or gaps in the observations.

Based on A24, the present study expands our scope to investigate decade-long periodicity in the historical optical LCs of 203 blazars in the Second Catalog of Hard Sources (2FHL) published by the *Fermi*-Large Area Telescope (LAT) collaboration (Ackermann et al. 2016). This investigation also assesses the impact of data gaps on our periodicity analysis. By evaluating the potential for these gaps to create artefacts in the LCs, we mitigate the risk of misinterpreting observational gaps as genuine periodic signals, enhancing our findings' robustness.

This work is organized as follows: Section 2 discusses access to historical optical data and data processing. Section 3 introduces the final sources for analysis using a general gap study on blazar-type data. The methodology is described in Section 4, followed by the results and discussion of the analysis and gap study in Section 5. The conclusions are given in Section 6.

## 2 DATA

We analyse optical data in our work primarily because optical telescopes have extensively monitored numerous blazars throughout the 20th century. In addition to that, observations in other wavelengths do not offer the century-long observation required for our objectives.

We search for long-term oscillation in blazars starting from the 2FHL catalogue, which helps us keep our initial sample size manageable. For these 203 blazars, we collect optical observations from various open-access data bases.

### 2.1 Archival optical data

Digital Access to a Sky Century at Harvard (DASCH; Grindlay et al. 2009) provides public access to a historical optical archive of numerous blazars monitored throughout the 20th century. By scanning photographic plates collected over a century from telescopes worldwide, DASCH facilitates new studies in time-domain astronomy, particularly enabling investigations into searching for decade-scale periodicity in blazar emissions.

Given that the magnitude data were digitized from photographic plates, there are inherent quality issues with the scanned data points. To mitigate these issues, we utilize data points with no quality flags and errors less than ±0.4 mag, as recommended.[1]

### 2.2 Complementary data bases

To complement the DASCH observations, we use data from other optical surveys, which provide *V*-band data on our sources of interest; Archives of Photographic PLates for Astronomical USE (APPLAUSE; Tuvikene et al. 2014),[2] the Katzman Automatic Imaging Telescope (KAIT; Filippenko et al. 2001),[3] the Catalina Sky Survey (CSS; Drake et al. 2009),[4] All-Sky Automated Survey for Supernovae (ASAS-SN; Shappee et al. 2014; Kochanek et al. 2017),[5] American Association of Variable Star Observers (AAVSO)[6] International Database, and Zwicky Transient Facility (ZTF; Masci et al. 2019).[7] These data bases extend the temporal coverage to 2024.

### 2.3 Data processing

The photometry data from DASCH, AAVSO, ASAS-SN, CSS, and ZTF are compatible without needing any offset correction, so their LCs can be combined directly, as demonstrated by A24. A24 used constant magnitude sources, like V∗ W Vir, to verify the compatibility of different data bases. We also found that V∗ W Vir's magnitude is in calibration with the APPLAUSE data base. Therefore, we can combine LCs from these data bases without any adjustments.

## 3 SOURCE SELECTION

We found data for 11 blazars on the DASCH data base out of the 203 blazars included in the 2FHL *Fermi*-LAT catalogue. Excluding PG 1553+113 as it was examined in our earlier work (see A24), we combine the DASCH data with that from the complementary data bases for our sources of interest, which are listed in Table 1 with their respective 2FHL source names, Right ascension (RA) and Declination (Dec.) coordinates, and redshifts (Ackermann et al. 2016).

We analyse the observed data in 28-d intervals. This binning period allows us to focus on variations that extend beyond the short-term, intraday, or week-long fluctuations typically observed in blazars. By adopting this approach, we can better identify the longer term patterns for assessing decadal-scale periodicity in blazar emissions. The binned LCs of the tabulated sources after filtering outliers using Tukey's fence method (Tukey et al. 1977) are shown in Fig. A1.

### 3.1 General gap study

Noticing the non-uniform sampling and gaps in the aforementioned LCs, we explore the impact of non-uniform sampling and gaps on periodicity study to further constrain the sample sources for our analysis. For that purpose, we simulate $N_{sim} = 10^6$ red-noise LCs with the power-law slope ($\alpha$) 1.5 and amplitude ($A$) 0.45, serving as toy model estimations for blazar power spectral density (PSD). 28-d binned data points going from 1900 to 2024 are generated for each simulated LC. These artificial data represent ideal and regular sampling. To mimic uneven sampling, 40 per cent of data are randomly removed, and to introduce regularly occurring gaps in the observations, an additional 45 per cent of data are removed in three regular intervals. Errors for each data point on each LC are introduced from a normal distribution $\mathcal{N}(0, 0.2)$. This process results in LCs that are randomly sampled, featuring three regular gaps and sampling, similar to the real ones, of 15 per cent. We then do a Generalized Lomb–Scargle Periodogram (GLSP) analysis on those LCs and calculate the significance of the highest peak in each of them against a red-noise signal. The result of this is shown in Fig. 1. The density plot reported in the top panel shows the distribution of observed periods, corresponding to uniformly distributed peak frequencies along with some edge effects and we do not notice a preference for any particular period due to the gaps. The histogram on the right shows the distribution of significances observed for peak periods in each LC; here, we notice a skewed Gaussian distribution with a mean significance of $\sigma_{mean} = 1.43$. The lopsided distribution hints at an increase in calculated significance due to gaps.

---

[1] http://dasch.rc.fas.harvard.edu/database.php#AFLAGS_ext
[2] https://www.plate-archive.org/applause/
[3] https://w.astro.berkeley.edu/bait/kait.html
[4] http://nesssi.cacr.caltech.edu/DataRelease/
[5] http://www.astronomy.ohio-state.edu/asassn/index.shtml
[6] http://https://www.aavso.org/data-download/
[7] https://www.ztf.caltech.edu/

**Table 1.** Periodicity and trend study candidates with their 2FHL names, RA and Dec. coordinates and redshifts.

| Association name | 2FHL source name | RA (J2000) | Dec. (J2000) | Redshift ($z$) |
| --- | --- | --- | --- | --- |
| PKS 0754+100 | J0756.8+0955 | 07 56 50.9 | +09 55 12 | 0.266 |
| TXS 1055+567 | J1058.5+5625 | 10 58 35.0 | +56 25 34 | 0.143 |
| MKN 421 | J1104.4+3812 | 11 04 28.8 | +38 12 26 | 0.031 |
| MKN 180 | J1136.5+7009 | 11 36 34.1 | +70 09 43 | 0.045 |
| M 87 | J1230.8+1225 | 12 30 49.0 | +12 25 54 | 0.004 |
| PG 1246+586 | J1248.1+5820 | 12 48 11.3 | +58 20 24 | – |
| PKS 1424+240 | J1427.0+2348 | 14 27 01.9 | +23 48 10 | 0.604 |
| PKS 1440−389 | J1443.9−3909 | 14 43 59.3 | −39 09 45 | 0.065 |
| AP Librae | J1517.7−2421 | 15 17 43.0 | −24 21 16 | 0.048 |
| MKN 501 | J1653.9+3945 | 16 53 54.2 | +39 45 14 | 0.034 |

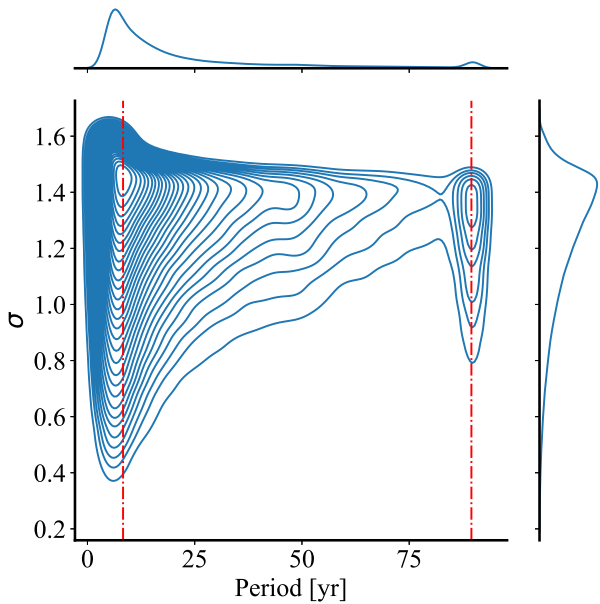

**Figure 1.** Distribution of observed peak periods and significances in the periodogram of simulated red-noise LCs with 85 per cent data removed. The top histogram is similar to a uniform distribution of frequencies except for the edge effects. We see this edge effect as an enhancement in the 8.3 and 89.5 yr periods, represented by the dash-dotted lines.

Additionally, we repeat the study for different fractions of sampling (such as 10 per cent, 20 per cent, and 25 per cent) and find that for a sampling of $\geq 15$ per cent, there is no preference for false periodicity, but a decrease in their significance with increased sampling is noticed. Therefore, we select the sources with at least 15 per cent sampling for our analysis.

### 3.2 Candidate sources

Finally, we select candidates for a periodicity study from the sources meeting the criterion defined in Section 3.1. The final selected sources are AP Librae (with 15.4 per cent sampling), M 87 (45.7 per cent), MKN 180 (15 per cent), MKN 421 (18.1 per cent), MKN 501 (23.3 per cent), and PG 1246+586 (15.9 per cent).

The LCs of the sources selected and discarded from further analysis using this criterion are shown in Fig. A1.

## 4 METHODOLOGY

This section discusses the methodology we use for the QPO analysis and significance estimation.

### 4.1 Periodicity estimation

The LCs of selected blazars are non-uniformly sampled, requiring periodicity estimation techniques that consider such time series data. GLSP (VanderPlas 2018) analysis and Weighted Wavelet Z-transform (WWZ; Foster 1996) are best suited to evaluate periodicity in such LCs, and both are widely used periodicity search methods in astronomy. In both methods, we search for periods in the range of 5 to 70 yr following A24 for all blazars in our final sample.

#### 4.1.1 Generalized Lomb–Scargle Periodogram

GLSP relates to the Fourier transform methods of periodicity analysis and the statistical method of least-squares, so it has a unique position for the periodicity study of unevenly sampled time series data (VanderPlas 2018). Fourier transform tells the relative amplitude of the frequencies present in the data, and the least-square method of periodicity analysis is flexible for unevenly sampled time series data. Following A24, we use the Astropy Collaboration (2022) implementation of GLSP on our data set.

#### 4.1.2 Weighted wavelet Z-transform

WWZ is a robust method employed to discern periodic signals within blazar LCs in A24. By employing a Morlet wavelet, WWZ allows for a time-frequency decomposition of the LC, revealing regions of significant power corresponding to periodic features. This method incorporates weights into the wavelet transform, which enhances sensitivity to specific frequency bands and time intervals and facilitates the detection of periodic signals amidst noise and variability. WWZ is famously useful for data having irregular sampling and noise contamination. This method identifies periodic signals with varying strengths and durations (e.g. Foster 1996). It is also important to consider the edge effects in the wavelet analysis, often represented by the cone of influence (COI) in WWZ plots. This is a region in the WWZ scalogram where the edge effects become significant, and the presence of a particular frequency becomes less reliable due to the decrease in the number of data points in the Morlet wavelet. In this work, we use the WWZ implementation of Khider et al. (2020; PYLEOCLIM).[8]

---

[8] https://github.com/LinkedEarth/Pyleoclim_util

**Table 2.** GLSP and Wavelet analysis results, their significance, and $R^2$ values of sinusoidal fits.

| Blazar name | GLSP (yr) | S/N | $R^2$ | WWZ (yr) | S/N | $R^2$ |
|---|---|---|---|---|---|---|
| AP Librae | 14.3±0.7 | PL: 2.3$\sigma$<br>BPL: 2.3$\sigma$ | 0.270 | 14.3±0.6 | PL: 2.7$\sigma$<br>BPL: 3.0$\sigma$ | 0.271 |
| M 87 | 45.4±1.1 | PL: 0.8$\sigma$<br>BPL: 0.7$\sigma$ | 0.260 | 49.0±7.0 | PL: 2.3$\sigma$<br>BPL: 2.7$\sigma$ | 0.290 |
| MKN 180 | 13.6±2.2 | PL: 1.1$\sigma$<br>BPL: 1.1$\sigma$ | 0.030 | 12.1±0.8 | PL: 1.5$\sigma$<br>BPL: 1.6$\sigma$ | 0.060 |
| MKN 421 | 47.6±9.5 | PL: 2.3$\sigma$<br>BPL: 2.4$\sigma$ | 0.230 | 54.2±8.8 | PL: 2.7$\sigma$<br>BPL: 2.8$\sigma$ | 0.230 |
| MKN 501 | 57.7±8.5 | PL: 1.5$\sigma$<br>BPL: 1.5$\sigma$ | 0.350 | 53.8±13.5 | PL: 2.1$\sigma$<br>BPL: 2.2$\sigma$ | 0.370 |
| PG 1246+586 | 36.4±5.2 | PL: 2.0$\sigma$<br>BPL: 2.2$\sigma$ | 0.090 | 31.7±3.1 | PL: 1.2$\sigma$<br>BPL: 1.8$\sigma$ | 0.150 |
|  | 13.1±0.3 | PL: 1.9$\sigma$<br>BPL: 2.1$\sigma$ | 0.126 | 12.9±0.6 | PL: 1.5$\sigma$<br>BPL: 2.1$\sigma$ | 0.128 |

**Table 3.** PL and BPL parameters for the sample blazars.

| | PL | | BPL | | |
|---|---|---|---|---|---|
| Blazar | $A$ | $\alpha$ | $A$ | $\nu_{bend}$ | $\alpha$ |
| AP Librae | 0.0090±0.0020 | 0.89±0.09 | 0.12±0.01 | 0.130±0.020 | 3.45±0.32 |
| M 87 | 0.0012±0.0001 | 0.91±0.04 | 0.11±0.02 | 0.030±0.010 | 1.13±0.08 |
| MKN 180 | 0.0040±0.0020 | 0.71±0.05 | 0.06±0.01 | 0.100±0.060 | 1.26±0.23 |
| MKN 421 | 0.0080±0.0010 | 0.74±0.03 | 0.29±0.07 | 0.020±0.010 | 1.02±0.08 |
| MKN 501 | 0.0015±0.0002 | 1.12±0.23 | 0.44±0.02 | 0.012±0.003 | 1.31±0.03 |
| PG 1246+586 | 0.0320±0.0020 | 0.89±0.08 | 0.38±0.02 | 0.020±0.010 | 0.85±0.05 |

### 4.2 Significance levels

Periodicity searches face challenges due to noise, which appears as erratic brightness fluctuations termed red noise (Papadakis & Lawrence 1993; Vaughan et al. 2003; Zhu & Xue 2016). We assess the significance of periods present in time series data for a complete periodicity analysis.

For that purpose, we simulate $10^6$ artificial LCs to determine the significance of detected periods and estimate the likelihood of false positives. These artificial LCs, generated following the technique of Emmanoulopoulos, McHardy & Papadakis (2013) using the PYTHON implementation by Foreman-Mackey et al. (2013), possess the same PSD, probability density function, and sampling characteristics as the real LCs. We consider two PSD models, power law (PL) and bending power law (BPL), to generate artificial LCs; the former provides a general prescription of red noise, and the latter provides a more realistic model of blazars' variability across all time-scales (e.g. Chakraborty & Rieger 2020). Subsequently, we analyse the resulting periodograms to determine the confidence levels of their peaks, calculated based on percentiles of the power for each period bin in the periodograms.

### 4.3 Power spectral density estimation

Traditionally, noise is categorized based on the power-law index $\alpha$ of the PSD equation $A * \nu^{-\alpha}$, where $\nu$ represents frequency (yr$^{-1}$), and $A$ denotes normalization (Rieger 2019). Additionally, for the BPL, we utilize the formula from Chakraborty & Rieger (2020):

$$P(\nu) = A \left(1 + \left\{\frac{\nu}{\nu_b}\right\}^{\alpha}\right)^{-1}, \quad (1)$$

where $A$ stands for normalization, $\alpha$ represents the spectral index, and $\nu_b$ signifies the bending frequency.

We estimate the parameters of each PSD model ($A$ and $\alpha$ for PL, and $A$, $\alpha$, and $\nu_b$ for BPL) for the six sources using maximum likelihood (ML) and Markov chain Monte Carlo simulations (MCMC; Foreman-Mackey et al. 2013).

## 5 RESULTS AND DISCUSSIONS

We present the results of our GLSP and Wavelet analyses on the six candidate blazars in Table 2. In both types of periodicity analyses, we calculate the significance of each of the observed periods against red noise following both PL and BPL methods (Table 3 shows the PSD fit parameters). As initial selection criteria, we take $2\sigma$ as a threshold in at least two out of four reported significances. We also take the fraction of the time series inside the COI in the scalogram plot as a measure of the quality of detection; we set a limit of $\geq 30$ per cent (see Fig. A2).

Following these selection criteria, the periods observed on the M 87 and MKN 180 LCs are considered non-detections. In the case of $\sim$ 13-yr period of MKN 180, none of the calculated significances are above $2\sigma$ (see Table 2). And for the observed period of $\sim$ 45 yr of M 87, < 20 per cent of the LC time-scale is within the COI (see Fig. A2). For the remaining four sources, we notice at least two $\geq 2\sigma$ significance, and the dominant period inside the COI spans at least 30 per cent of the time series length.

### 5.1 Candidates

We consider the periods of the four sources (AP Librae, MKN 421, MKN 501, and PG 1246+586) as potential detections and further scrutinize the observed periodicity. We fit a sinusoidal signal on the LCs and take the goodness-of-fit ($R^2$) value as a relative measure of the quality of observed periods. As we do not expect the periodicity in blazar emission to follow a perfect sinusoid, we refrain from using these values as an absolute measure of the periodicity and rather compare the $R^2$ among the results. These values are reported in Table 2. Furthermore, we do a phase fold of the LCs with the observed periods, making visualization of a periodic signal easier, especially when the data are uneven and with gaps. Below, we discuss these additional studies in the periodicity for each blazar.

### 5.1.1 *AP Librae*

For the source AP Librae, we show the LC, periodicity analysis results, and the 14.3-yr phase-folded LC in Fig. 2. We see similar $R^2$ values for GLSP and WWZ periods. When compared with other sources, the goodness of fit is also high. The reported significances are highest for AP Librae compared to other sources, and we do not notice any gaps on the phase-folded LC (see Fig. 2).

### 5.1.2 *MKN 421*

For this source, we observe a $\sim 55$ yr period on the LC with $> 2\sigma$ significance with the same $R^2$ value of 0.23 for GLSP and WWZ periods. The observed period falls at the upper range of the search period, where the impact of red noise is more pronounced, and we notice only a limited fraction of the signal ($\sim 40$ per cent) inside the COI. The phase-folded LC, however, does not show data gaps.

### 5.1.3 *MKN 501*

We observe a $\sim 55$ yr period for MKN 501 with $> 2\sigma$ significance on only the wavelet analysis. The $R^2$ values are similar for GLSP and WWZ periods and are highest among the sources. However, we notice a limited signal ($\sim 40$ per cent) inside the COI, and the peak of this period is also not well defined in the wavelet PSD plot (see Fig. A2). We also do not notice gaps in the phase-folded LC data for this source.

### 5.1.4 *PG 1246+586*

For this source, we observe two simultaneous periods of $\sim 35$ yr and $\sim 13$ yr. However, the $R^2$ values for the GLSP and WWZ periods are the lowest among our candidate sources, indicating the poorest periodic behaviour among our sources of interest. The 36-yr phase-folded LC shows a contiguous gap in the signal (see Fig. A3), suggesting that the data gap influences this period, but we do not notice such gaps on the 13 yr phase-folded LC.

## 5.2 Gap study for candidate sources

The gaps in the LCs could have an impact on the observed periods of AP Librae, MKN 421, MKN 501, and PG 1246+586. This fact demands further analysis of the impact of the gaps in the data. This section discusses the effect of the gaps in the significance estimation and the false periodicities due to the uneven sampling for each of our sources. For this analysis, we only report our candidates' GLSP period(s) and PL significance for simplicity.

We conduct an analogous analysis to Section 3.1 using our sample blazar's binned LCs. We use the PL fit parameters from Table 3 for each of the sources. We maintain the same sampling and

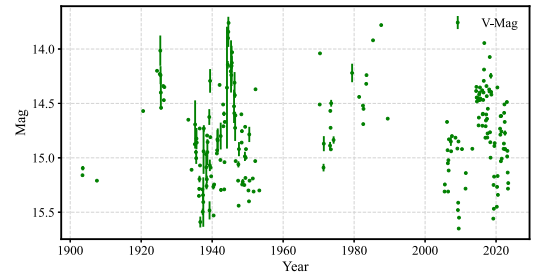

(a) 28 days binned LC combined from DASCH and complementary databases.

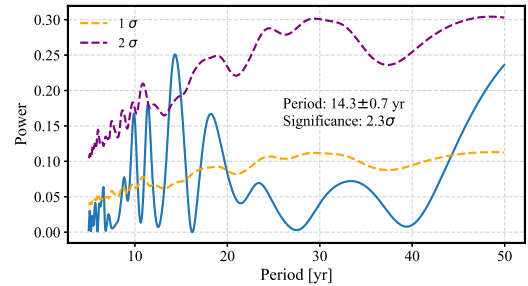

(b) GLSP periodogram with significance levels. The peak period of $14.3 \pm 0.7$ yr has a PL significance of 2.3 $\sigma$.

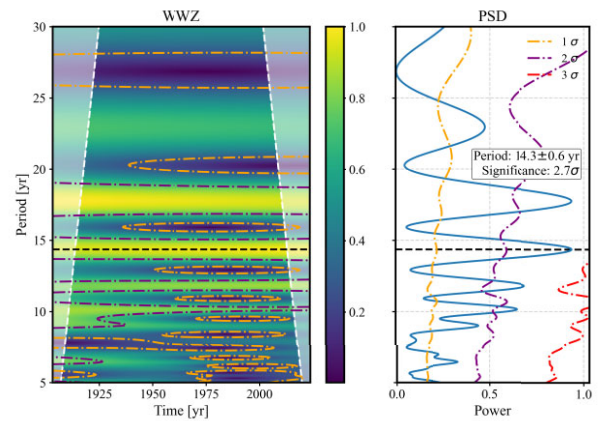

(c) WWZ scalogram and PSD showing the highest peak at $14.3 \pm 0.6$ yr with $2.7\sigma$ PL significance.

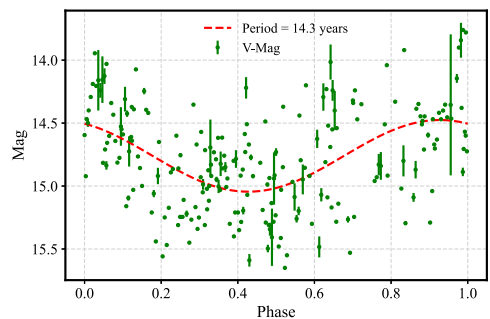

(d) 14.3 yr phase-folded LC with a sinusoidal fit. The $R^2$ of the sine fit is 0.27.

**Figure 2.** AP Librae LC and periodicity analysis results.

observation uncertainties for each simulated signal as the original LC. For instance, for each AP Librae-like red-noise LC, we keep the same temporal sampling as the original binned LC and simulate the magnitudes with PL fit parameters $A = 0.009$ and $\alpha = 0.002$ and on top of that, we assign magnitude errors by resampling the errors from the original LC. The results of this analysis are presented in a period-significance distribution plot. Where the distribution shows the observed peak periods and significances in the periodogram of blazar-like red-noise LCs. The vertical red-dotted lines represent enhancement locations, and the red+ symbol indicates the observed GLSP period and PL significance from Table 2. We discuss the results in detail for each source below.

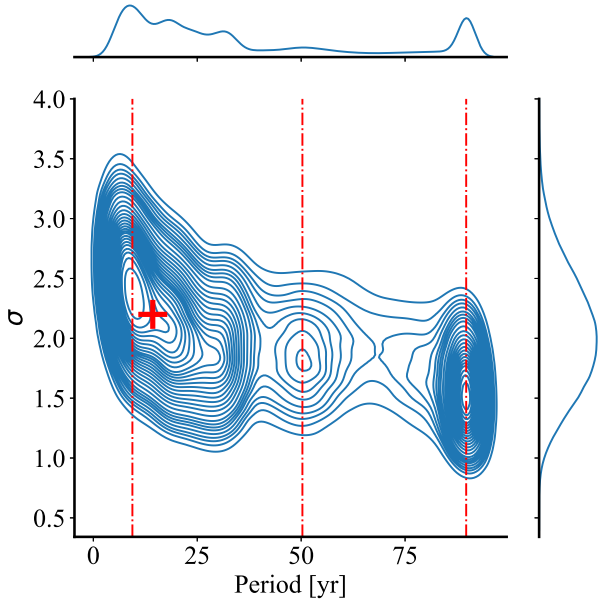

**Figure 3.** For AP Librae-like red-noise LCs, enhancements are at 9.4, 50.3, and 89.7 yr with the observed period (+) of 14.3 yr (2.2$\sigma$) inside the enhancement due to the gap.

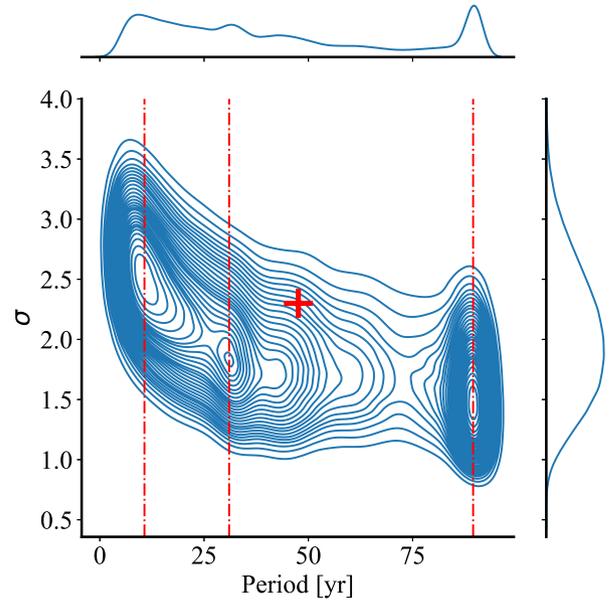

**Figure 4.** For MKN 421-like red-noise LCs, enhancements are at 10.7, 31.0, and 89.5 yr with the observed period (+) of 47.6 yr (2.3$\sigma$) away from the enhancement due to gap.

### 5.2.1 AP Librae

The period-significance distribution for AP Librae-like LCs reveals notable spikes in the period density plot around 10 and 30 yr (see the top panel of Fig. 3), in contrast to the general gap study (see the top panel of Fig. 1). Additionally, we observe an overall elevation in significance values for all detected periods, especially for periods <30 yr (see the right panel of Fig. 3). The period we observe (14.3 yr) lies within the enhancement. Additionally, the significance density plot is centered at 1.9$\sigma$ (> $\sigma_{mean}$ from Section 3.1), suggesting the significance value we report (2.2$\sigma$) can be caused by the nature of the gaps.

### 5.2.2 MKN 421

The distribution for MKN 421-like LCs reveals notable spikes in the period density plot around 10 and 30 yr (see the top panel of Fig. 4). Furthermore, we observe an overall elevation in significance values for all detected periods, similar as for AP Librae. The main plot shows that the observed period (47.6 yr) lies outside the enhancement region, and the significance density plot on the right is centred at 1.6$\sigma$ (> $\sigma_{mean}$). Our result suggests the period and its significance value we report (2.2$\sigma$) is less likely due to the gaps.

### 5.2.3 MKN 501

For the MKN 501-like LCs, the period density function (top panel of Fig. 5) shows a slight enhancement around 57 yr. Although this enhancement is relatively minor, the significance distribution (right panel) has a mean of 1.5$\sigma$, which puts our observed period's significance of 1.5$\sigma$ within the possibility of spurious detection. We conclude that this period could be due to the gaps in the observation.

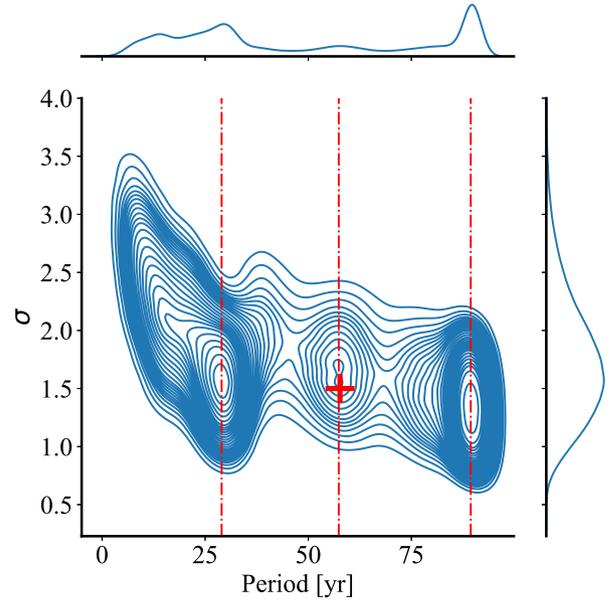

**Figure 5.** For MKN 501-like red-noise LCs, enhancements are at 29.0, 57.4, and 89.4 yr with the observed period (+) of 57.7 yr (1.5$\sigma$) inside the enhancement due to the gap.

### 5.2.4 PG 1246+586

For this source, the period density plot on the top panel of Fig. 6 shows enhancements for periods less than 40 yr. Both the observed periods fall within this enhancement region. The significance density detected period, which can be approximated using the formula:

$$p_{\text{global}} = 1 - (1 - p_{\text{local}})^N, \quad (2)$$

where $N$ represents the trial factor. The expression calculates this trial factor:

$$N = P \times B, \quad (3)$$

where $P$ denotes the number of independent periods (frequencies) in each periodogram, and $B$ is the number of blazars in our sample. The value of $P$ is estimated through Monte Carlo simulations, utilizing the algorithm described by Peñil et al. (2022) and employing $10^6$ simulated light curves generated using the method by Timmer & Koenig (1995). Through an exploration of values of $P$, we aim to achieve the best fit to the experimental relationship of local-global significance, selecting a value of $P$ to correct significance levels of $\geq 2\sigma$ obtained in our analysis. With a limit of $P \leq 100$ to balance computational efficiency and resolution, we find $P = 43$.

Given that we analysed six blazars ($B = 6$), the number of independent periods determines the trial factor, resulting in a total trial factor of $N = 258$. Applying this factor to equation (2), the global significance for a significance level of $2\sigma$ is approximately zero. This means the period we observe in this work are statistically non-significant.

### 5.5 False alarm probability

False alarm probability (FAP) is the probability that a blazar-like, red-noise LC produces a peak period with a given local significance. We use $2\sigma$ as a representative value of the local significances, as discussed in Section 5 for FAP calculation. We generate $10^6$ LCs based on two PSD models: PL and BPL, employing the parameters estimated in Section 4.3 while maintaining the same sampling interval as the 28 d binned LCs of our sources. We determine the following false detection rates for our sample blazars.

Our analysis reveals approximately 36.7 per cent and 27.6 per cent false detection rates for PL and BPL, respectively, for AP Librae. These values correspond to a global significance of $\sim 0.3\sigma - 0.6\sigma$, which are again statistically non-significant.

### 5.6 Light curve reconstruction and prediction

Even though the period observed in the LC of AP Librae can not be conclusively attributed to genuine physical processes, we perform a sinusoidal fit on the LC and predict the next high and low emissions. While the underlying periodic behaviour (if genuine) may not strictly follow a sinusoidal curve, extending the fit and predicting the future emissions might provide insights into the underlying cyclic pattern. We present the sinusoidal fit and the predictions in Fig. 7.

## 6 SUMMARY AND CONCLUSIONS

We searched historical optical data for the blazars included in the 2FHL *Fermi*-LAT catalogue. As a result, we found data for 11 of the blazars in DASCH and other data bases. We exclude PG 1553+113 from our study as it was studied in our previous work (see A24). From the remaining 10 blazars, we excluded PKS 0754+100, PKS 1424+240, PKS 1440–389, and TXS 1055+567 because their data did not meet our requirement in terms of the percentage of gaps in their LCs. Out of the remaining six sources, we found hints of QPOs in four sources: AP Librae, MKN 421, MKN 501, and PG 1246+586. We found a QPO of 14.3 yr with a local significance of $2.2\sigma$ for AP Librae, 47.6 yr ($2.3\sigma$) for MKN 421, 57.7 yr ($1.5\sigma$) for plot has a mean of $1.8\sigma$, putting our observed periods and reported significance into question.

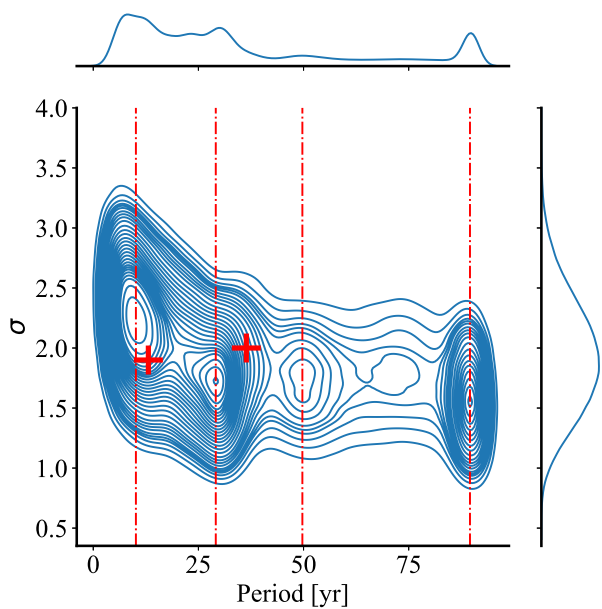

**Figure 6.** For PG 1246+586-like red-noise LCs, enhancements are at 10.1, 29.1, 49.7, and 89.6 yr with both observed periods (+) of 13.1 yr ($1.9\sigma$) and 36.4 yr ($2.0\sigma$) close to the enhancement due to the gap.

Given a multitude of quality issues with the observed periods, such as the unreliable significance values, the gap in phase-folded LC, and the low $R^2$ values of the fits of both periods, we omit PG 1246+586 from further analyses.

From the analysis of the sources of interest, only the $\sim$50-yr period of MKN 421 is likely not an artefact of the gaps. More importantly, these findings corroborate the argument that gaps in time series data can yield spurious periods with higher significance, suggesting caution in interpreting the observed period in the LCs.

### 5.3 Number of cycles criterion

In addition to the gap study of each source, another important criterion in blazar QPO is the number of observed cycles. A minimum of three to five complete cycles is typically required to reliably distinguish true periodicity from red noise in long-term LCs (Vaughan et al. 2016). Out of the four sources discussed in Section 5.1, the $\sim$47 yr period of MKN 421 (2.7 cycles), $\sim$57 yr period of MKN 501 (2.0 cycles), and the $\sim$36 yr period of PG 1246+586 (2.9 cycles) do not meet this criterion. Given that these periodicities span less than three full cycles over the observational baseline, we consider these periodicities as non-detections. Although the phase plots may appear compelling (see Fig. A3), the limited number of cycles makes their interpretation statistically uncertain.

Therefore, the only QPO in candidate in consideration for further analysis is $\sim$14.3 yr period of AP Librae (8.4 cycles).

### 5.4 Global significance

The significance obtained from our analysis methods needs adjustment to account for the look-elsewhere effect, as outlined by Gross & Vitells (2010). This correction yields the global significance of the MKN 501, and double periods of 13.1 yr ($1.9\sigma$) and 36.4 yr ($2.0\sigma$) for PG 1246+586.

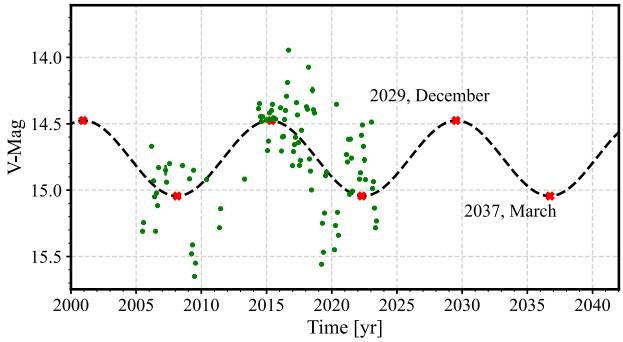

**Figure 7.** Prediction of the optical emission from the sinusoidal reconstruction of AP Librae. The next maxima and minima will occur in 2029 December and 2037 March, respectively.

In addition to that, our gap-impact analysis on blazar-like red-noise signals showed that the observed periods of three sources (AP Librae, MKN 501, and PG 1246+586) are likely an artefact of the nature of data sampling, specifically the gaps in the observation. Further employing the number of cycles criterion, we only analysed AP Librae for global significance. The QPO we reported is not statistically significant after correcting for the look elsewhere effect. The global significance was $\sim 0\sigma$'s, and the FAPs were $\sim 0.3\sigma - 0.6\sigma$.

With more continuous monitoring of blazars in optical and other wavelengths, we could determine whether the low-significance QPOs we found are real. If genuine, this would allow us to constrain the properties of the central engine and test binary black hole hypotheses. Furthermore, decade-long periodicities could indirectly identify binary systems, potentially identifying merger events for future gravitational wave detectors.

## ACKNOWLEDGEMENTS

We want to thank all the observatories from which we used data. The DASCH project at Harvard is grateful for partial support from NSF grants AST-0407380, AST-0909073, and AST-1313370. Funding for APPLAUSE has been provided by DFG (German Research Foundation, Grant), Leibniz Institute for Astrophysics Potsdam (AIP), Dr Remeis Sternwarte Bamberg (University Nuernberg/Erlangen), the Hamburger Sternwarte (University of Hamburg), and Tartu Observatory. Plate material also has been made available from Thüringer Landessternwarte Tautenburg, and from the archives of the Vatican Observatory. We thank the Las Cumbres Observatory and its staff for their continuing support of the ASAS-SN project. ASAS-SN is supported by the Gordon and Betty Moore Foundation through grant GBMF5490 to the Ohio State University, and NSF grants AST-1515927 and AST-1908570. Development of ASAS-SN has been supported by NSF grant AST-0908816, the Mt. Cuba Astronomical Foundation, the Center for Cosmology and Astroparticle Physics at the Ohio State University, the Chinese Academy of Sciences South America Center for Astronomy (CAS-SACA), the Villum Foundation, and George Skestos. The AAVSO data base: Kafka, S., 2021, Observations from the AAVSO International Database, https://www.aavso.org. The National Aeronautics and Space Administration funds the CSS survey under Grant No. NNG05GF22G issued through the Science Mission Directorate Near-Earth Objects Observations Program. The Catalina Real-Time Transient Survey is supported by the U.S. National Science Foundation under grants AST-0909182 and AST-1313422. Based on observations obtained with the Samuel Oschin Telescope 48-inch and the 60-inch Telescope at the Palomar Observatory as part of the Zwicky Transient Facility project. ZTF is supported by the National Science Foundation under Grant No. AST-2034437 and a collaboration including Caltech, IPAC, the Weizmann Institute for Science, The Oskar Klein Center at Stockholm University, the University of Maryland, Deutsches Elektronen-Synchrotron and Humboldt University, the TANGO Consortium of Taiwan, the University of Wisconsin at Milwaukee, Trinity College Dublin, Lawrence Livermore National Laboratories, and IN2P3, France. Operations are conducted by COO, IPAC, and UW.

MA and PP acknowledge funding from NASA under contracts 80NSSC24K0635, 80NSSC23K1040, 80NSSC23K0294, and 80NSSC22K1578.

AD is thankful for the support received from the Proyecto PID2021-126536OA-I00, funded by MCIN/AEI/ 10.13039/501100011033.

This work was supported by the European Research Council, ERC Starting grant MessMapp, SB Principal Investigator, under contract no. 949555, and by the German Science Foundation DFG, research grant 'Relativistic Jets in Active Galaxies' (FOR 5195, grant No. 443220636).

## DATA AVAILABILITY

All the data used in this work are publicly available or available on request to the person responsible for the corresponding observatory/facility.

# APPENDIX

This section reports the LCs of the blazars with historical data and the results of the WWZ for those selected according to the gap criterion defined in Section 3.2.

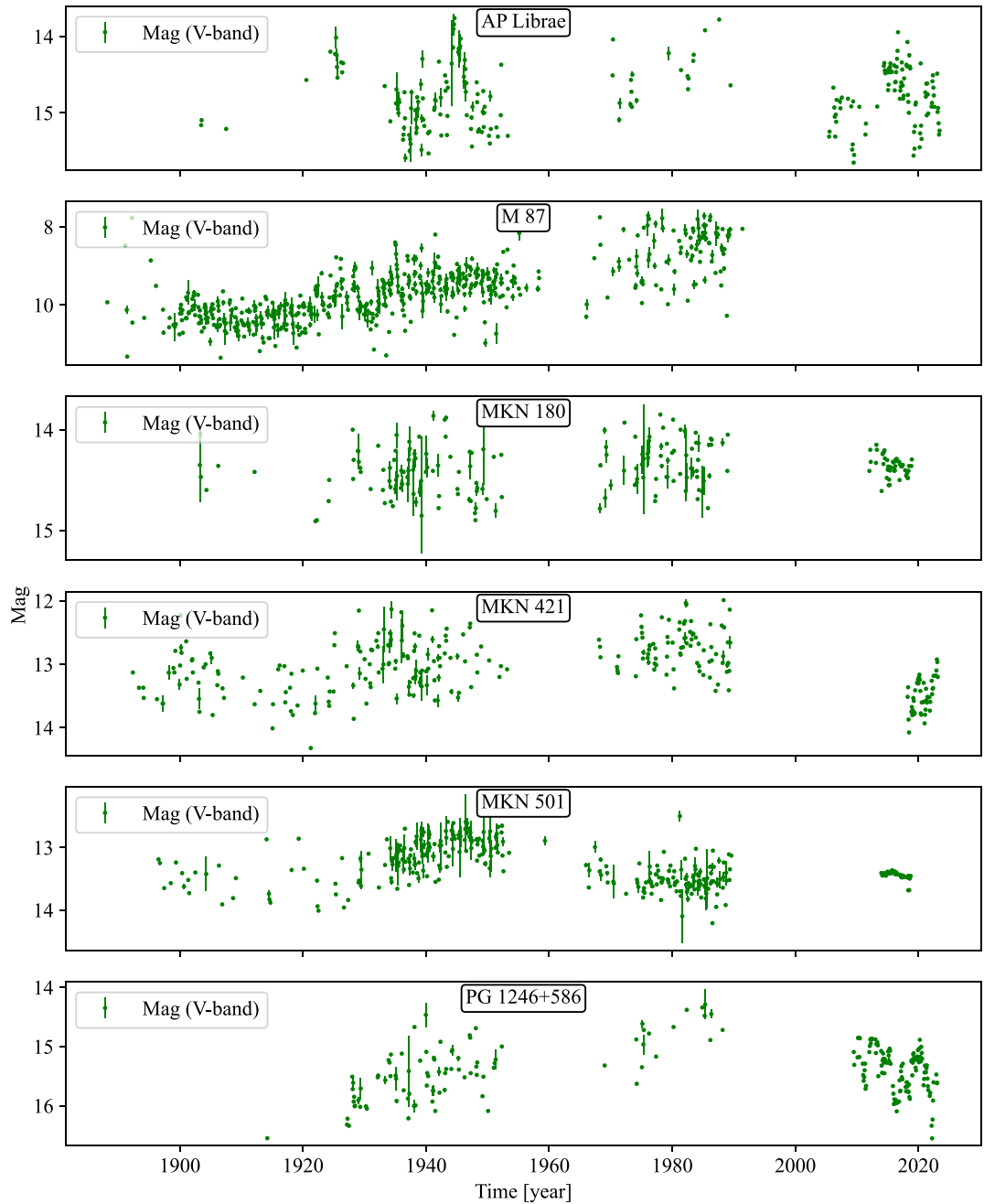

**Figure A1.** 28 d binned light curves of blazars selected for periodicity study.

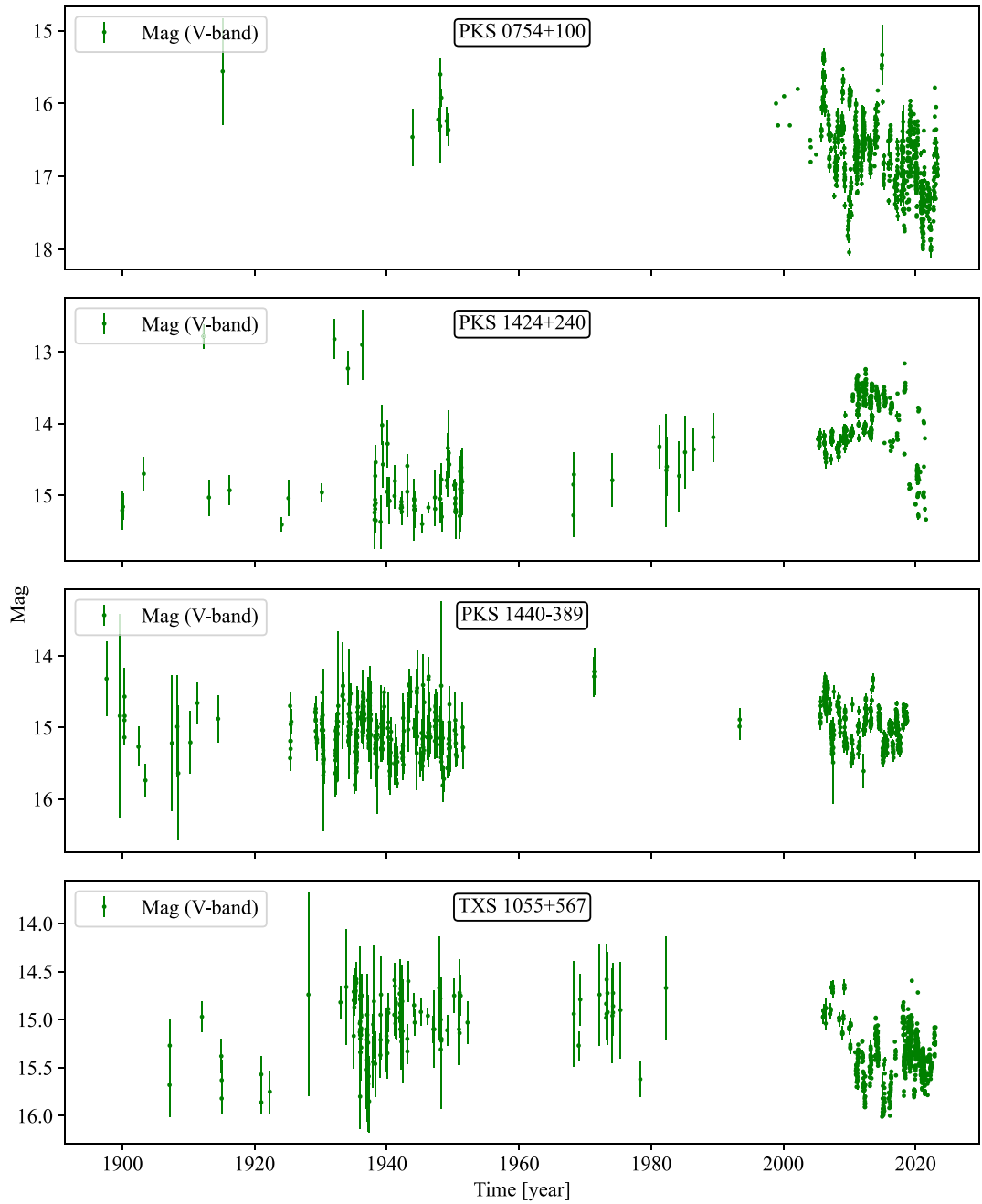

**Figure A1.** *Continue*

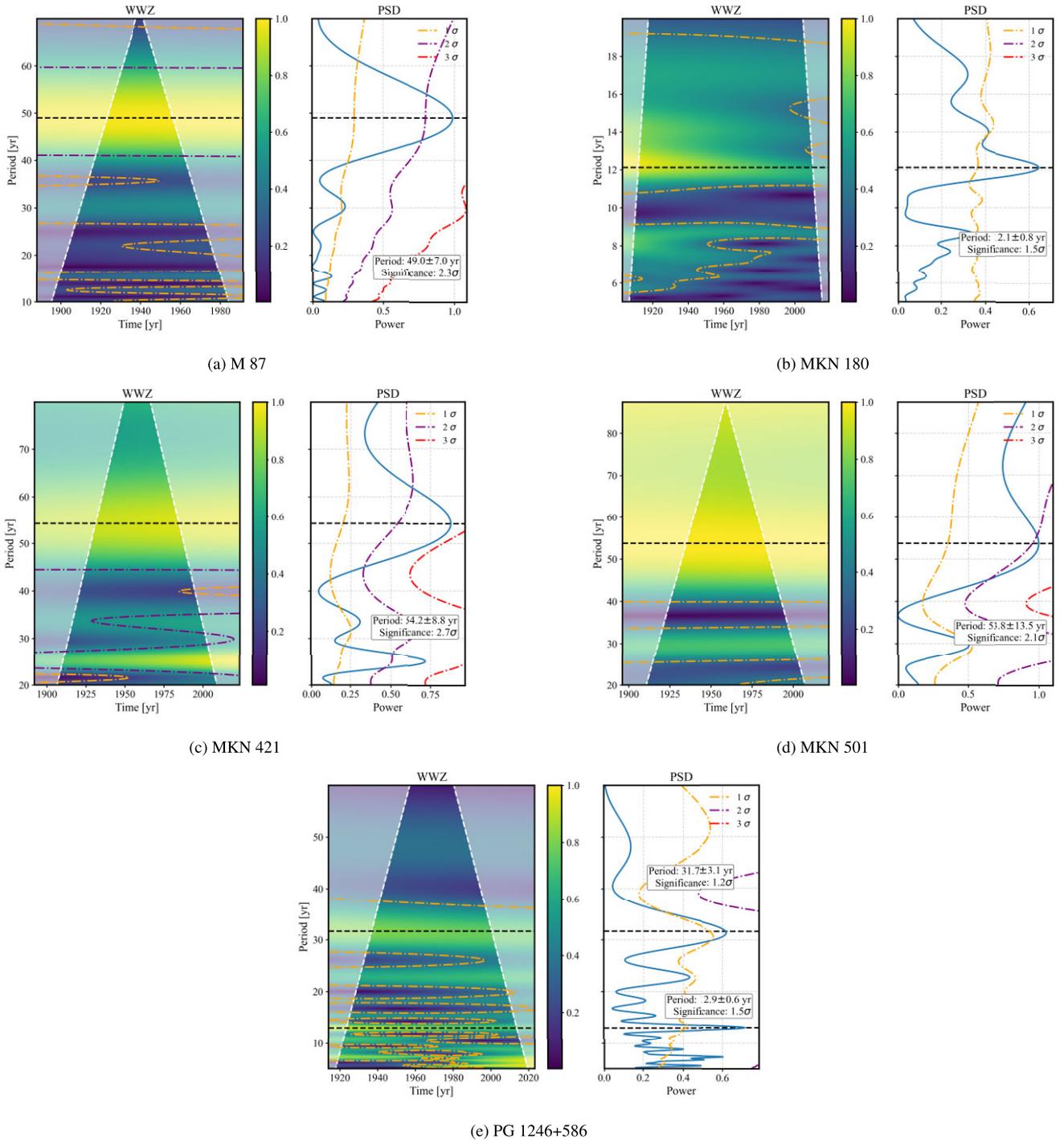

**Figure A2.** WWZ scalograms and PSD plot for the rest of the sources from Table 2. The shaded light region shows the region outside the COI, the dashed lines represent prominent peaks, and dash–dot lines of different colours represent different significance and contour levels.

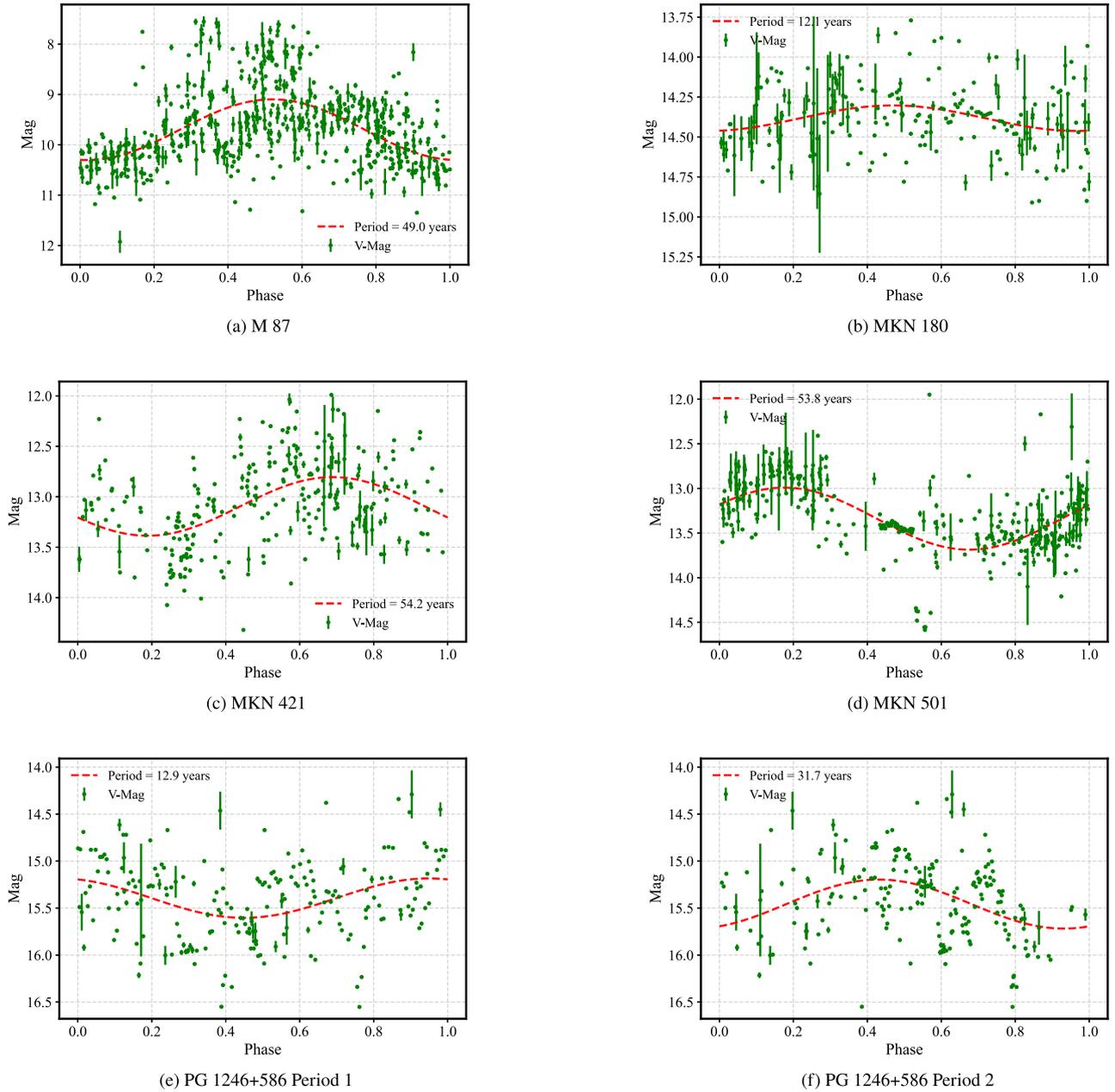

**Figure A3.** Phase plots for the rest of the sources from Table 2. The two simultaneous periods of PG 1246+586 are shown in panels (e) and (f).

This paper has been typeset from a T<sub>E</sub>X/L<sup>A</sup>T<sub>E</sub>X file prepared by the author.